\documentclass{article}
\usepackage{amssymb}

\input{tcilatex}
\begin{document}

\title{The Area of a Rough Black Hole}
\author{John D. Barrow \\
DAMTP, Centre for Mathematical Sciences,\\
University of Cambridge,\\
Wilberforce Road, Cambridge CB3 0WA\\
United Kingdom\ \ \ \ \ \ \ \ }
\maketitle
\date{}

\begin{abstract}
We investigate the consequences for the black hole area of \ introducing
fractal structure for the horizon geometry. We create a three-dimensional
spherical analogue of a 'Koch Snowflake' using a infinite diminishing
hierarchy of touching spheres around the Schwarzschild event horizon. We can
create a fractal structure for the horizon with finite volume and infinite
(or finite) area. This is a toy model for the possible effects of quantum
gravitational spacetime foam, with significant implications for assessments
of the entropy of black holes and the universe, which is generally larger
than in standard picture of black hole structure and thermodynamics,
potentially by very considerable factors. The entropy of the observable
universe today becomes $S\approx 10^{120(1+\Delta /2)}$, where $0\leq \Delta
\leq 1$, with $\Delta =0$ for a smooth spacetime structure and $\Delta =1$
for the most intricate. The Hawking lifetime of black holes is also reduced.
\end{abstract}

\section{ Introduction}

\ Mathematicians are familiar with constructions like the Koch snowflake 
\cite{koch} in which a two-dimensional self-similar object, constructed
iteratively, can possess finite area and infinite perimeter. In three
dimensions, the Sierpinski Gasket \cite{sier} and the Menger Sponge \cite%
{men} have a analogous properties, with finite volume and infinite surface
area. These features are perfectly consistent with the isoperimetric
theorems that relate surface area, $A,$ with enclosed volume$,V$, for a
3-dimensional body by the inequality $A^{3}\geq 36\pi V^{2}$, with equality
for the sphere. Since we are being bombarded with animations and pictures of
the Covid-19 virus as a sphere with a number of attachments leading off its
surface to increase its surface area and provide links to latch on to other
cells, we question whether, at the quantum gravitational level, space and
black hole surfaces might be like that, with intricate structure down to
arbitrarily small scales (or to a cut-off scale of order the Planck length),
leading to an increase over the expected surface area. The surface area of a
black hole is a key feature that gives its entropy and information content.
It obeys a 'second law', or Area Theorem, subject to energy conditions, that
requires it to be non-decreasing. Bekenstein \cite{bek} and Hawking \cite%
{swh} discovered many of these crucial classical properties whose
significance for physics is wider than the study of black holes. Hawking 
\cite{swh2} showed that they are not mere coincidences or analogies with
thermodynamic laws, as was once thought, but deep consequences of the
quantum structure of a black hole: black holes are black bodies.

\ \ In section II, we outline a fractal extension of the surface structure
of the static, spherically symmetric Schwarzschild black hole and determine
conditions needed for the volume to remain finite while the surface area
tends to infinity in the limit of increasing intricacy on arbitrarily small
scales. We highlight a number of consequences for the entropy and Hawking
lifetime of black holes and for assessments of the entropy of the observable
universe. In section III we discuss the physical bases for this type of
horizon structure and in section IV we discuss our results and their
limitations.

\section{Black hole with an intricate surface}

We will construct a fractal horizon surface starting from a Schwarzschild
black hole of mass $M$ and radius $R_{g}=2GM/c^{2}$ by attaching some number
of smaller spheres to touch its outer surface with yet smaller spheres
touching the surfaces of those spheres, and so on. The original Koch
snowflake boundary in 2-dimensions is made of a crenellated structure of
increasingly small triangles whose sides have the middle third converted
into the base of a new equilateral triangle with sides that are three times
smaller: our boundary will be composed of surfaces of hierarchically
smaller, touching, spheres. Suppose that each step to smaller scale
intricacy leads to the attachment of $N$ spheres of radius $\lambda $ times
smaller than the sphere to the sphere to which they are attached
tangentially. Therefore, the hierarchy of radii is just $r_{n+1}=\lambda
r_{n}$ over $N$ steps, where $r_{0}=R_{g}$, is the Schwarzschild radius.

If we allow this process of adding smaller spheres to touch the surface,
then the total volume of the black hole \ after an infinite number of steps, 
$V_{\infty }$, will be

\begin{equation}
V_{\infty }=\dsum\limits_{n=0}^{\infty }N^{n}\frac{4\pi }{3}\left( \lambda
^{n}R_{g}\right) ^{3}=\frac{4\pi R_{g}^{3}}{3}\dsum\limits_{n=0}^{\infty
}(N\lambda ^{3})^{n}.  \label{0}
\end{equation}%
This is a finite convergent so long as $N\lambda ^{3}<1$. This ensures the
geometric series on the right-hand side of eq.(\ref{0}) converges. In that
case, the $N\rightarrow \infty $ limit is

\begin{equation}
V_{\infty }=\frac{4\pi R_{g}^{3}}{3(1-N\lambda ^{3})}\ >\frac{4\pi R_{g}^{3}%
}{3\ }  \label{1}
\end{equation}%
Therefore, the volume of the extended fractal black hole is finite under
these conditions.

Similarly, the total surface area after an infinite number steps, $A_{\infty
}$, is

\begin{equation}
A_{\infty }=\dsum\limits_{n=0}^{\infty }N^{n}4\pi (\lambda
^{n}R_{g})^{2}=4\pi R_{g}^{2}\dsum\limits_{n=0}^{\infty }(N\lambda
^{2})^{n}>4\pi R_{g}^{2}.  \label{2}
\end{equation}%
Since we want the surface area to diverge in the limit we require $N\lambda
^{2}>1.$ When $N\lambda ^{2}<1,$ the area converges to

\begin{equation}
A_{\infty }=\ \frac{4\pi R_{g}^{2}}{1-N\lambda ^{2}}.  \label{2a}
\end{equation}

Hence, the volume will be finite but the surface area \ will be infinite if

\begin{equation}
\lambda ^{-2}<N<\lambda ^{-3}.  \label{3}
\end{equation}

The divergence of the surface area in the limit, if it is achieved rather
than the sum being cut off at some small finite radius, renders the black
hole entropy infinite, and probably meaningless as a physical indicator.
However, if it converges to a finite limit, or has a cut-off length, the
area is again always greater than the spherical Schwarzschild surface area,
eq. (\ref{2}), as we might expect from the classical area theorem.

There is another restriction to consider: the number of spheres that will
fit around the sphere of the previous iteration. If we just consider a
two-dimensional slice and fit as many circles of radius $r$ around a bigger
circle of radius $R_{g}$, then the circle that passes through the centres of
all the smaller circles that touch the larger one has radius $R_{g}+r.$ The
maximum number of circles we can pack in the first level of the hierarchy is
given by $Nr=\pi (R+r)$; so, if $r=\lambda R_{g}$, as above, we have the
bound

\begin{equation}
N\leq \pi (\lambda ^{-1}+1).  \label{4}
\end{equation}%
The true bound will be 3-dimensional, but this is slice estimate is
indicative and concordant with eq. (\ref{3}).

The surface area, $A_{g}=4\pi R_{g}^{2}$, of a Schwarzschild black hole
determines its entropy, $S=A_{g}c^{3}/4G\hbar \approx A_{g}/A_{pl}$, where $%
A_{pl}$ is the Planck area: the entropy is the number of Planck areas in the
horizon area. Thus, we see that with intricate horizon structure, if the
thermodynamic interpretation of the area still holds as its fundamental
thermodynamic basis might suggest, then the entropy of the black hole can be
much larger than the standard Schwarzschild value as it is arising in a
quantum gravitational extension of general relativity and its usually
assumed spacetime structure.\ Thus we cannot assume that the usual
principles for black holes (no hair, entropy bound etc) will hold in
unchanged form. The increased value is what we could expect from an Area
Theorem,\emph{\ }$dA/dt\geq 0$\emph{,} with the increased complexity and
information needed to describe the horizon structure, leading to a higher
entropy. Likewise, in this context the evaporating quantum black hole with
the increased area will lead to more rapid evaporation by Stefan's law.
There will be a shorter lifetime before the black hole explodes, since the
luminosity is proportional to $A_{g}T_{g}^{4}$, and $T_{g}\varpropto M^{-1}$
is the black hole temperature. If the area increases by a scaling $%
A_{g}\rightarrow \alpha A_{g}$, via eq. (\ref{2}), with $\alpha \geq 1$,
then the black hole's Hawking lifetime, $t_{bh}$, falls as $t_{bh}\varpropto
M^{3}/\alpha ^{2}$ as the intricacy, $\alpha $, increases$.$ If there is no
upper bound on $\alpha $, then primordial black holes will explode very
quickly and may leave no direct explosive remnants today.

In a more general scheme, where the surface of the black hole is a pure
fractal we know that the surface area will vary as the radius to a power%
\emph{\ }$R^{2+\Delta }$\emph{,} where $0\leq \Delta \leq 1,$\ with $\Delta
=0$\ corresponding to the simplest horizon structure, and $\Delta =1$\ to
the most intricate, where it behaves from an \ information perspective as if
it possessed one geometric dimension higher. Thus, from this perspective the
black hole entropy would vary as $S\approx (A/A_{pl})\approx
(A_{g}/A_{pl})^{(2+\Delta )/2}$ . For an application of this formula to the
observable universe inside the particle horizon today we take $A_{g}\approx
(ct_{0})^{2}$, with the present cosmic age $t_{0}\approx 10^{17}s$, so we
have

\begin{equation}
S_{u}\approx (10^{17}/10^{-43})^{(2+\Delta )}\approx 10^{120(1+\Delta /2)}\ ,
\label{5}
\end{equation}%
and it ranges between the usual $10^{120}$ with smooth spacetime structure
and $10^{180}$ with the most fractalised. Likewise, the entropy of a fractal
black hole possesses a similar enormous range of possible entropy values for
a given mass.

\section{Physical Motivations}

Our toy example is just intended to show that near the scale where quantum
gravity effects impinge, the surface area of a black hole can greatly exceed 
$4\pi R_{g}^{2}$\ because of intricate small-scale structure of fractal
type. This will occur for any external intricacy with a Hausdorff dimension
exceeding 2. In effect, the 2-dimensional geometrical surface behaves as
through it has more than two dimensions and approaches the behaviour of a
3-dimensional surface in the limit of maximum intricacy, showing that it has
the information content and intricacy of a geometrical volume\footnote{\emph{%
This way of increasing effective area is widespread in the natural world,
for example, if \ you feel the the crinkled surface of an elephant's skin it
must scale faster than the square of any measure of its size span (as the
elementary biology texts wrongly assume) to allow for more efficient cooling
than occurs if it is simply proportional to the standard geometric area.}}.
Although we know almost nothing about spacetime structure on scales within a
few orders of magnitude of the Planck scale, where we might expect to find
these complexities in the geometry, the first suggestions of a spacetime
foam structure were suggested by Wheeler \cite{JW} as a model of spacetime
structure on the Planck scale, see also \cite{swh3, hooft}. On larger
scales, this model has become one of three paradigms for observational
testing on larger astronomical scales. Recently, the strongest limit have
been found using\ Espresso \cite{cooke}\ at the VLT through its effect on
images and the profile stability of the FeII metal-line velocity, $v,$.
Under the assumption that the effects are proportional to $(E/E_{pl})^{a}$,
where $E_{pl}$\ is the Planck energy, the effects on $\Delta v/c$\ are
proportional to $(1+z)^{-1-a}$, with $1/2\leq a\leq 1,$\ where the light
source is at redshift $z=2.34,\ $about $5.8Gpc$\ away from us. The random
walk model has $a=1/2,$ the holographic model has $a=2/3,$\ while Wheeler's
model has\emph{\ }$a=1;$\emph{\ }but Wheeler's model, unlike the other two,
produces no cumulative effects over the spacetime path from source to
detector and so is not open to investigation by observing light from
high-redshift astronomical sources. The limits from the first two scenarios
are that $a\geq 0.625,$so they exclude some random-walk models. If photons
take discrete random walk steps en route to us then those steps must be at
least $10^{13.2}$\ Planck lengths $(10^{-29.8}cm)$\ in size. This is a 3-4
order of magnitude improvement over earlier bounds on spacetime foam from
observations of distant quasars by the Chandra x-ray Satellite and the Fermi
Gamma-ray Space Telescope, coupled with ground-based gamma-ray observations
from the Very Energetic Radiation Imaging Telescope Array (VERITAS) \cite%
{chandra}. They claim that spacetime must be uniform down to distances of
order $10^{-16}$\ $cm$\ in order not to diffuse incoming light from the
quasars and degrade image quality by unacceptable levels, but this is still
far above the $10^{-33}$\ $cm$\ Planck length scale at which quantum
gravitational foam might be expected.

\ Other scenarios with a foam-like picture of spacetime microstructure have
been studied in some detail, for example the spinfoam theory \cite{spin}.
The most generic property of quantum theories, including a quantum gravity
theory that is yet to be found, which motivates our simple fractal-like
structure for the black hole horizon, is that of the fractal nature of
quantum paths. This was first mentioned in Snyder \cite{snyd} and has been
reviewed in ref. \cite{cas}, and references therein.

Feynman and Hibbs \cite{feyn} pointed out that the 'typical path of a
quantum particle is highly irregular on a small scale.. in other words [it
is] non-differentiable', and illustrate the structure pictorially. Many
other authors observe similarities between Brownian and quantum-mechanical
motions (see, for example, Nelson \cite{nel} and references herein).
Similarly, the dimension of the quantum path was also discussed as the
dimension of a non-differentiable path in quantum field theories by Kraemmer
et al \cite{krae}. Later, after the introduction of the fractal terminology
by\emph{\ }Mandelbrot in 1972 \cite{man}, Abbott and Wise \cite{wise} later
calculated that the fractal (Hausdorff) dimension of the quantum path in one
dimension is 2, i.e. maximal, so with the information content of an area.
The reason for this fractal behaviour is fundamental and this is why we
expect it to be possible on very small Planck length scales in 3-dimensional
space around the rough black hole horizon. The reason for its domination of
quantum paths is the Heisenberg Uncertainty Principle (HUP). As a particle
becomes more localized in a region $\Delta r$\ its momentum\emph{\ }becomes
of order $1/\Delta r$\ and its motion becomes more erratic. Abbott and Wise 
\cite{wise} show that when the step-lengths are much larger than the quantum
wavelength of the particle the the Hausdorff dimension, $D$,\ approaches 1
but when the step-lengths are smaller than the quantum wavelength $D$\
approaches 2, with the information content of a geometric volume, thus
showing the fractal character on arbitrary small scales that we have
exploited in our simple model above. In between these limiting cases the
fractal dimension varies rapidly but exceeds 1. Theories of generalised
Hagedorn type with a continuously rising spectrum of mass states of the form 
$g(m)\varpropto (m/m_{0})^{\beta },$\ for constants $m_{0}>0$\ and $\beta
>1, $\ display structure on arbitrarily small scales as the quantum
wavelength of the mass states declines when $m\rightarrow \infty $, \cite{BH}%
, and again such scenarios are well suited to create microscopic fractal
structure in combination with quantum random motion.

An interesting extension of these calculations is to replace the HUP by its
extension when gravitational forces are included. The uncertainty in
position $\Delta r$\ and momentum $\Delta p$\ in one dimension is then

\begin{equation}
\Delta r\text{ }\simeq \frac{1}{\Delta p}+\lambda l_{pl}^{2}\Delta p.
\label{6}
\end{equation}%
\ The first term on the right-hand side of eq. (\ref{6}) is the term in the
usual HUP. The second term reflects the horizon fluctuation, $\Lambda
R_{g}\simeq \Delta M_{bh}\simeq \Delta p,$\ where $\lambda $\ is some
geometrical constant of order unity) and the intricate structure we have
argued should appear near the horizon on length scales close to the Planck
length.

\section{\protect\bigskip\ Discussion}

The deficiencies of our model are clear. We do not create the fractal
substructures by any single quantum gravity model (because there is no such
standard model). However, we have discussed some particular theories for
spacetime foam and the non-differentiable character of quantum particle
paths, with and without the presence of gravity, to motivate our scenario.
The fact that it can rest on such simple general physical principle adds to
its plausibility\ and makes the hypothesis worthy of further exploration.
There are many other ways we could have constructed a 'snowflake' structure
of the horizon on arbitrarily small scales but we chose the simplest toy
model. Using this, we have explored the general effect on the event horizon
areas of black holes and inside the cosmological particle horizon. Similar
effects can alter the assessment of the 'likelihood' of the whole universe
as the black hole entropy formula is often used to assess the gravitational
entropy of the universe, by asking for the entropy of the largest black hole
that could fit into it \cite{pen, BT, JDB}, or the number of Planck volumes
that will fit inside the particle horizon. We have seen how these black-hole
and cosmological entropies can even be infinite if there is no small-scale
cut-off to the intricacies. This is often assumed but is not proven. The
laboratory analogue studies of black hole horizons might also be able to
investigate these changes to the horizon intricacy directly \cite{barc, unr}%
. We also discussed new observational probes of the scale of any spacetime
foam structure using its effects on astronomical images and spectral lines. 
\emph{Since this paper was first posted a number of detailed studies have
appeared examining the observational consequences of the fractalised area
for dark matter and Hubble tension \cite{bas, sar, sar2} and for aspects of
black hole thermodynamics \cite{abreu, bas2} and the spontaneous
'wrinklification' of AdS black holes \cite{McInn}.}

This is a fascinating, albeit very model dependent although now we are only
able to probe far larger length scales than we expect fractal effects to
make the horizon of a black hole 'fuzzy', they are welcome steps towards
closing the link between theory and observation.

Acknowledgement. The author is supported by the Science and Technology
Funding Council of the UK (STFC) and thanks S. Cotsakis, E. Vagenas, M. D\c{a%
}browski and S. Vagnozzi for helpful inputs and references\emph{.}

\end{document}